# B20-MnSi films grown on Si(100) substrates with magnetic skyrmion signature

Zichao Li[1, 2*], Ye Yuan[1, 3], René Hübner[1], Viktor Begeza[1, 4], Thomas Naumann[1,], Lars Rebohle[1], Olav Hellwig[1, 5], Manfred Helm[1,4], Kornelius Nielsch[2, 4, 6], Slawomir Prucnal[1], Shengqiang Zhou[1*]

[1]Helmholtz-Zentrum Dresden-Rossendorf, Institute of Ion Beam Physics and Materials Research, Bautzner Landstrasse 400, D-01328 Dresden, Germany

[2]Institute of Materials Science, Technische Universität Dresden, 01069 Dresden, Germany

[3]Songshan Lake Materials Laboratory, Dongguan, Guangdong 523808, People's Republic of China

[4]Institute of Applied Physics, Technische Universität Dresden, 01062 Dresden, Germany

[5]Institute of Physics, Chemnitz University of Technology, Chemnitz 09126, Germany

[6]Institute for Metallic Materials, IFW-Dresden, Dresden, 01069, Germany

* Corresponding author: Zichao Li and Shengqiang Zhou

E-mail address: zichao.li@hzdr.de and s.zhou@hzdr.de

Abstract: Magnetic skyrmions have been suggested as information carriers for future spintronic devices. As the first material with experimentally confirmed skyrmions, B20-type MnSi was the research focus for decades. Although B20-MnSi films have been successfully grown on Si(111) substrates, there is no report about B20-MnSi films on Si(100) substrates, which would be more preferred for practical applications. In this letter, we present the first preparation of B20-MnSi on Si(100) substrates. It is realized by sub-second solid-state reaction between Mn and Si via flash-lamp annealing at ambient pressure. The regrown layer shows an enhanced Curie temperature of 43 K compared with bulk B20-MnSi. The magnetic skyrmion signature is proved in our films by magnetic and transport measurements. The millisecond-range flash annealing provides a promising avenue for the fabrication of Si-based skyrmionic devices.

## 1. Introduction

In noncentrosymmetric magnetic materials, the Dzyaloshinskii Moriya (DM) interaction can cause spins to arrange non-collinearly [1-3], and together with the ferromagnetic exchange interaction, it results in a whirling domain structure called skyrmions [4-8]. Due to the small size and smaller motion energy for skyrmions, they can be used in new-generation integrated storage devices [9, 10]. Magnetic skyrmions were experimentally observed for the first time in bulk ferromagnetic B20-type MnSi [11]. Afterwards, a lot of effort has been devoted to the growth of B20-MnSi thin films [12-14]. Due to the small lattice mismatch, the films were mostly grown on Si(111) substrates [12-14]. For instance, B20-MnSi films on Si(111) were successfully prepared by molecular beam epitaxy (MBE) [13-15], by solid-state phase epitaxy (SPE) [12, 16], and magnetron sputtering [17]. In most of the cases, MnSi films grown on Si(111) show a Curie temperature of around 43 K, i.e. much higher than the 29.5 K reported for bulk MnSi. However, so far, there has been no report about the formation of B20-MnSi on Si(100) substrate, which is more preferred for spintronic devices due its compatibility with complementary metal-oxide-semiconductor (CMOS) technology. Note that the Si(100) wafer has a higher symmetry than the Si(111) one and its cutting and etching have been well developed [18]. It seems that the growth of B20-MnSi on Si(100) is not possible due to the large mismatch of 16% [19]. Indeed, the $MnSi_{1.7}$ phase is more easily grown on Si(100) substrates and captures researchers' attention due to its excellent thermoelectric properties [20, 21]. $MnSi_{1.7}$ films were prepared on Si(100) by SPE [22] and MBE [23]. Furthermore, Wu *et al.* and Kahwaji *et al.* found that B2-MnSi can be formed on Si(100) substrates [24, 25]. However, B2 MnSi cannot host skyrmions since it is centrosymmetric [24]. The absence of B20-MnSi on Si(100) substrates significantly limits the application of skyrmions in integrated spintronics devices. Therefore, it is desirable to search for different growth approaches for B20-MnSi on Si(100) substrates. Xie et al. have demonstrated the growth of $Mn_5Ge_3$ films on Ge(100) by millisecond solid-state reaction, between Mn and Ge via flash-lamp annealing [26]. Note that by normal solid-state reaction $Mn_5Ge_3$ is more easily formed on Ge(111) substrates [27]. This therefore suggests that a non-equilibrium annealing method could allow for the nucleation of a crystalline phase, which is otherwise not favourable under thermal equilibrium condition.

In this work, we use millisecond flash-lamp annealing induced solid-phase reaction to synthesize B20-MnSi films. We show that the nucleation of B20-MnSi on Si(100) substrates is possible, leading to a continuous polycrystalline thin film, although the MnSi$_{1.7}$ parasitic phase cannot be eliminated. The Curie temperature of the fabricated films, compared to bulk B20-MnSi, increases to around 43 K. The characteristic signature of magnetic skyrmions in our MnSi thin films is proved by magnetization and magneto-transport measurements [13]. The skyrmions can be stabilized from 1000 Oe to 7000 Oe at the whole temperature range below 43 K.

## 2. Experimental methods

To grow B20-MnSi films, we first deposited Mn films with thicknesses varying from 7 to 30 nm on Si(100) substrates by magnetron sputtering. Afterwards, millisecond flash-lamp annealing (FLA) with 20 ms pulse duration in a continuous N$_2$ flow was used to heat the sample and to trigger the solid-phase reaction between Mn and Si [28]. During FLA, a bank of capacitors is discharged over a field of flash lamps, producing a light pulse of high intensity. The sample surface will absorb the light and its temperature can reach more than 1000 K within the ms range. Since the pulse duration is only 20 ms, this process results in large heating and cooling rates. Such a non-equilibrium process leads to a very high phase formation speed, making it possible to realize non-equilibrium material phases, like B20-MnSi on Si(100), which is otherwise not possible under thermal equilibrium. The obtained Mn-Si films are around 14-60 nm thick. The topography of the regrown layers was observed by visible-light microscopy. XRD was performed at room temperature by using a Rigaku Smart Lab diffractometer with a Cu-target source. The measurements were done in Bragg-Brentano-geometry with a graphite secondary monochromator and a scintillator detector. Additionally, micro-Raman spectroscopy and transmission electron microscopy (TEM) were employed for microstructure analysis. The micro-Raman experiments were done using a Horiba micro-Raman system with the excitation wavelength of 532 nm and a spot size of 1 μm, and the signal was recorded with a liquid-nitrogen-cooled silicon CCD camera. Bright-field and high-resolution TEM images were recorded with an image-C$_s$-corrected Titan 80-300 microscope (FEI) operated at an accelerating voltage of 300 kV. High-angle annular dark-field scanning transmission electron microscopy (HAADF-STEM) imaging and spectrum imaging analysis based on energy-dispersive X-ray spectroscopy (EDXS) were performed with a Talos F200X microscope (FEI) operated at 200 kV to obtain the element composition. The magnetic properties of the films were measured by a superconducting quantum interference device equipped with a vibrating sample magnetometer (SQUID-VSM) with the field parallel (in-plane) to the films. The transport properties of MnSi films were investigated by a Lake Shore Hall measurement system. Magnetic-field (in-plane) dependent resistance was measured between 5 and 45 K using the van der Pauw geometry.

## 3. Results and Discussions

3.1. Phase identification and structure characterization

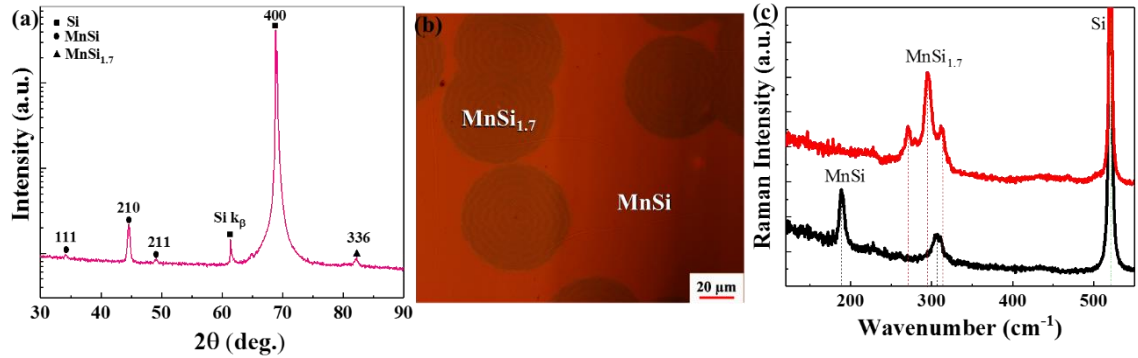

Figure 1. (a) XRD and (b) visible-light microscopy image of the sample annealed from a 30-nm-thick Mn layer on Si(100) substrate. The grey concentric circles are the phase of $MnSi_{1.7}$ and the featureless area is B20-MnSi. The scale bar is 20 µm. (c) Raman spectra obtained at different positions as shown in (b). For the darker concentric circles and the matrix in-between, the Raman spectra were measured for more than 6 positions. Each matrix position shows the same Raman signal as the black curve, pointing to the B20-MnSi phase. The spectrum of each circular area shows three peaks at around 300 cm$^{-1}$, consistent with the $MnSi_{1.7}$ phase.

We first focus on the sample grown from 30 nm Mn on Si(100). Figure 1(a) shows the XRD pattern of this sample. The (400) diffraction peaks of the Si substrate are at 69.5°. The MnSi(111), (210) and (211) Bragg peaks are observed at 34.2°, 44.6° and 49.2°, respectively. According to the powder PDF card (n. 01-081-0484) [23], the MnSi (210) peak is the strongest. Taking into account the intensity ratio between different peaks, the MnSi film grown on Si(100) is of polycrystalline nature. From the XRD pattern, we can confirm the formation of B20-MnSi on Si(100). The parasitic $MnSi_{1.7}$ phase always coexist with MnSi phase and its (336) Bragg peaks are observed at 82.2°. As shown by a representative visible-light microscopy image of the regrown layer [Fig. 1(b)], there are two different areas: one is composed of darker concentric circles with a diameter of around 60 µm and the other one is the matrix area in-between. To understand the phase formation for these two areas, the corresponding room-temperature Raman spectra are shown in Fig. 1(b). All presented Raman spectra show the peak at around 520.5 cm$^{-1}$ (dashed green lines), which corresponds to the transverse/longitudinal optical (TO/LO) phonon mode in the Si substrate [29]. Additionally, the matrix area shows two well-separated peaks at about 188 and 307 cm$^{-1}$ (dashed black lines), which correspond to the *E* or *T* Raman-active phonon modes in MnSi [30, 31]. In fully relaxed bulk MnSi, these two phonon modes are located at around 194 and 316 cm$^{-1}$ [30, 31]. The shift of the phonon mode positions toward lower wavenumbers indicates the existence of in-plane biaxial tensile strain. The Raman spectrum obtained from the area of the concentric circles exhibits three separated peaks at around 300±20 cm$^{-1}$ (dashed red lines). These peaks confirm the formation of the MnSi$_{1.7}$ phase. These three phonon modes should account for the slightly different neighbourhoods of the Mn–Si bonds in MnSi$_{1.7}$ [32]. Based on the Raman results, we can conclude that the circular and matrix areas exhibit the MnSi$_{1.7}$ and B20-MnSi phase, respectively. More specifically, MnSi$_{1.7}$ is embedded in the B20-MnSi matrix. By rough estimation from the visible-light microscopy images, the areal fraction of the MnSi phase comprises around 45%. A more reliable estimation of this value is given by comparing the saturation magnetization of our films with that of bulk MnSi in the next section. The precipitation of MnSi$_{1.7}$ in the shape of concentric circles is assumed to be caused by an inhomogeneous temperature distribution or compositional fluctuations [33]. We notice that the formation of MnSi$_{1.7}$ within the sample of around 10×10 mm$^2$ is random. Inevitably, some local regions have lower temperature or redundant Si, both favouring the nucleation of MnSi$_{1.7}$ [28]. However, more investigations are needed for a complete understanding and to control the ratio between MnSi and MnSi$_{1.7}$.

Figure 2(a) shows a representative cross-sectional bright-field TEM image of the FLA-treated 30-nm-thick Mn layer on Si(100) obtained from the matrix area indicated in Fig. 1(b). Below a 5-to-20-nm-thin amorphous surface film (light-grey with uniform brightness), there is an about 50-nm-thick polycrystalline film which is mainly composed of grains having a height equal to and a lateral extension larger than the film thickness (dark-gray appearance with slightly varying orientation contrast). To further characterize the phase structure of the continuous polycrystalline layer, high-resolution TEM imaging (Fig. 2(b)) combined with fast Fourier transform analysis (Fig. 2(c)) was performed. Taking the B20-type MnSi structure, the diffractogram in Fig. 2(c) can be described by a [0 $\bar{1}$ 0] zone axis pattern (in particular: normal to the growth plane [3 0 2] MnSi is parallel to [0 0 1] Si and in-plane [0 $\bar{1}$ 0] MnSi is parallel to [1 $\bar{1}$ 0] Si). The spatial arrangement of the elements Mn, Si, and O was characterized by EDXS-based analysis in scanning TEM mode and is shown in Fig. 2(d). In particular, Mn and Si are homogeneously distributed within the silicide layer, and the Mn:Si atomic ratio is determined to be close to 1:1. Above the MnSi layer, there is an oxide layer composed of Mn, Si and O. The small oxygen signal within the Mn silicide region is caused by TEM lamella side-wall oxidation during storage in air. For thicker TEM lamella positions, this O signal is even less pronounced. In summary, TEM analysis confirms the formation of a continuous polycrystalline B20-type MnSi film.

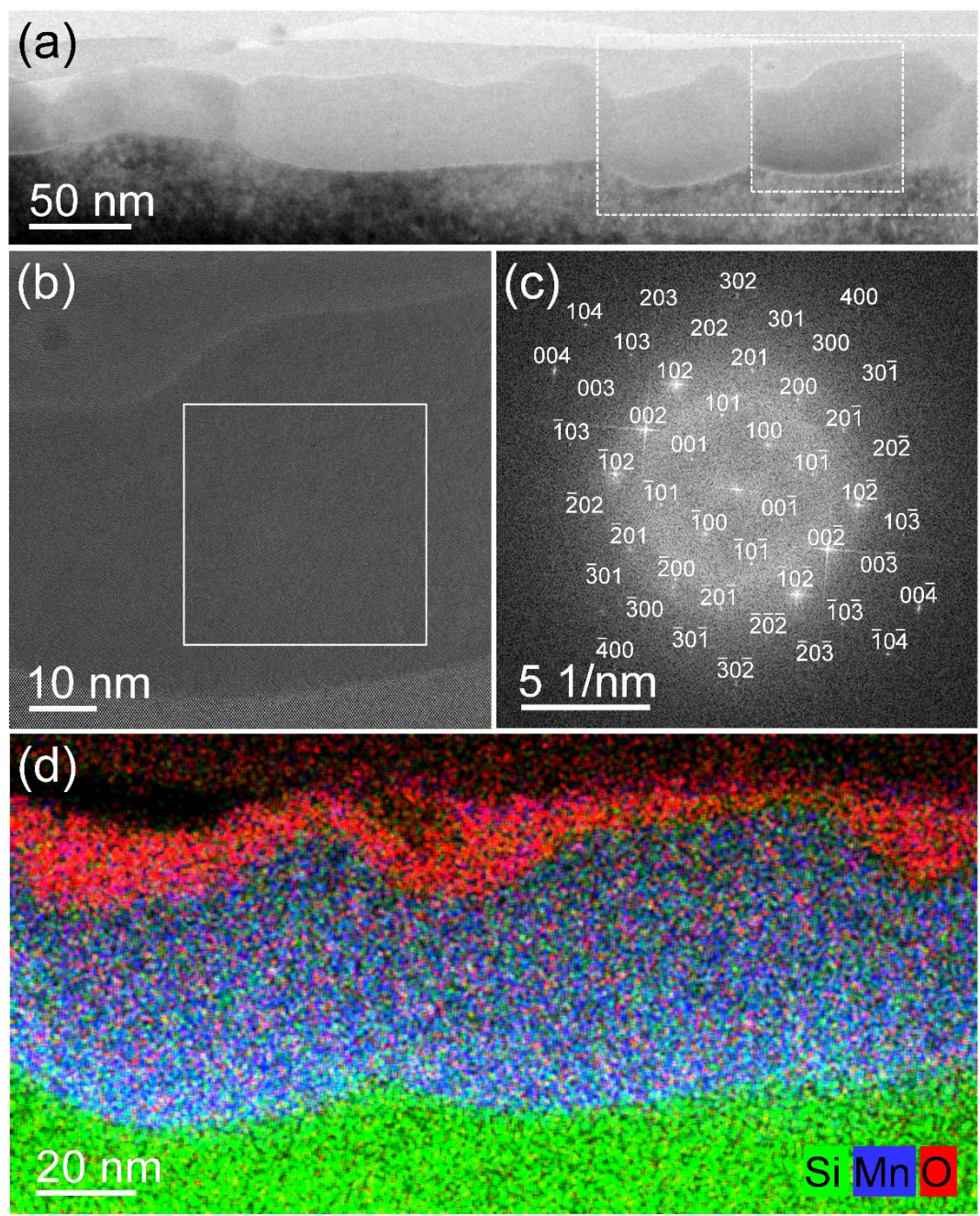

Figure 2. (a) Cross-sectional bright-field TEM image of the FLA-treated 30-nm-thick Mn layer on Si(100) substrate obtained from the matrix area shown in Fig. 1(a). (b) High-resolution TEM image of the area marked with a dashed white square in panel (a). (c) Fast Fourier transform of the region marked with a white square in panel (b) and indexed based on a MnSi [0 $\bar{1}$ 0] zone axis pattern. (d) Superimposed EDXS-based element distributions (blue: manganese, green: silicon, red: oxygen) for the area marked with a white dashed rectangle in panel (a), confirming the presence of a continuous B20-MnSi film.

## 3.2. Magnetic measurement and skyrmion signature

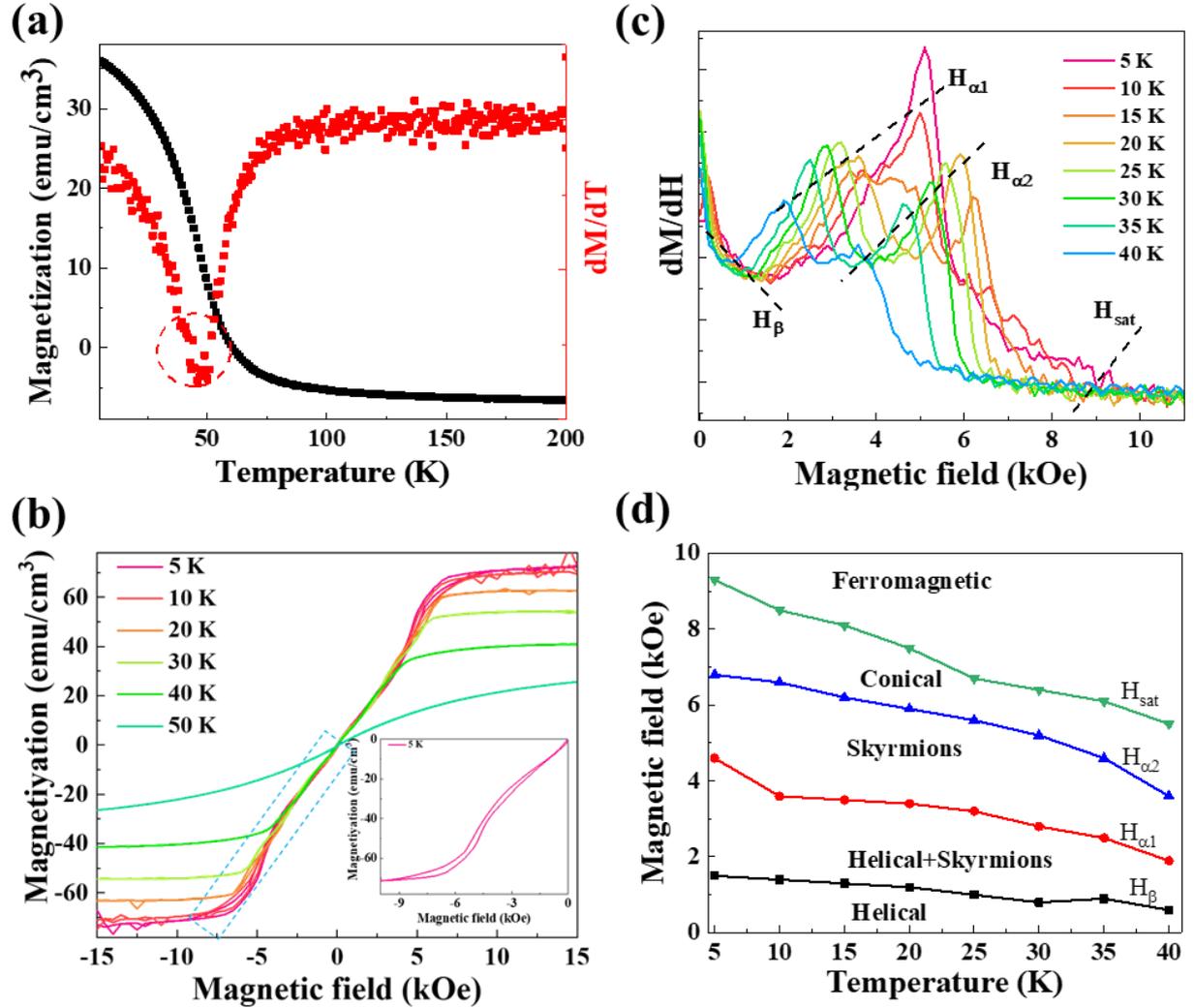

Figure 3. Magnetic properties of B20-MnSi made by solid-state reaction of a 30-nm-thick Mn layer on Si(100) substrate. (a) In-plane saturation magnetization (black squares) and the calculated derivative dM/dT (red squares). The valley of dM/dT at around 43 K indicates the Curie temperature. (b) In-plane M-H curves measured at different temperatures. (c) Static susceptibility in increasing field sweeps for the MnSi film at various temperatures. The black dashed lines in (c) show the shift of transition fields at each temperature. $H_{\alpha1}$ and $H_{\alpha2}$ bound the region of stable elliptic skyrmions. The dashed lines define the boundary for metastable helicoids $H_{\beta}$ and ferromagnetic $H_{sat}$. (d) Magnetic phase diagram. With increasing temperature, the mentioned magnetic phases formed at lower magnetic fields.

Figure 3(a) displays the temperature-dependent saturation magnetization of the MnSi alloy obtained from the 30-nm-thick Mn layer on Si(100) after FLA. The field was applied along the film surface namely, in-plane. As indicated by the minimum of the derivative dM/dT (red square) curve, the Curie temperature of our B20-MnSi film is 43 K and thus higher than that for bulk MnSi (29.5 K [15]). This Curie temperature enhancement is still under discussion and is very probably caused by strain or defects [34]. It is known that $MnSi_{1.7}$ is a weak itinerant magnet with a saturation magnetization of 0.012 $\mu_B$/Mn [35] compared with 0.43 $\mu_B$/Mn for B20-MnSi [36], which is why the measured magnetization must originate from the B20-MnSi phase. The in-plane MH curves in Fig. 3(b) show multi-hysteresis below the Curie temperature, which arises from the variety of different spin arrangements. To better observe the multi-hysteresis, the insert of Fig. 3(b) is an enlargement of the blue dashed rectangle. Our measurements also indicate that the transition between different magnetic structures occurs within the whole temperature range, not only near the Curie temperature as for bulk MnSi [11]. The saturation magnetization at 5 K is around 72 emu/cm$^3$. If we neglect the magnetization of MnSi$_{1.7}$ and compare with the saturation magnetization of 163 emu/cm$^3$ for bulk MnSi [11], the MnSi volume fraction in our sample is around 45%. This value is in reasonable agreement with that obtained by calculating areal fraction from Fig. 1(a).

The derivatives of the static magnetization dM/dH are often used to identify the magnetic phase boundaries or critical fields in many skyrmion materials [12, 13, 28]. In Fig. 3(c), the derived dM/dH curves of our films show four critical transition fields, named as $H_\beta$, $H_{\alpha 1}$, $H_{\alpha 2}$, and $H_{sat}$ with the dashed lines indicating their shift with temperature. Below $H_\beta$, the system stays at its ground helical state. The next apparent peaks at $H_{\alpha 1}$ and $H_{\alpha 2}$ mark the first-order transitions in and out of a new phase with a difference in topology, namely the so-called skyrmion phase [37]. According to this, the region between $H_\beta$ and $H_{\alpha 1}$ is a transition region, where the ground helical and the skyrmion phase coexist. Above $H_{\alpha 2}$, the skyrmion phase almost vanishes and the system evolves into a conical state, followed by another transition to field-polarized ferromagnetism above the critical field of $H_{sat}$. Based on the temperature dependency of the transition fields a magnetic phase diagram as shown in Fig. 3(d), can be constructed. With increasing the magnetic field, the Helical, Helical + Skyrmions, Skyrmions, Conical and Ferromagnetism phase emerge in this order [17]. It is in good agreement with other B20-MnSi thin films with in-plane skyrmions which are stabilized by the uniaxial magnetic anisotropy [38, 39]. It is interesting that our B20-MnSi film on Si(100) also hosts skyrmions over a wide temperature and magnetic-field range similar to what has been shown in literature [40, 41]. This enlarged skyrmion stability offers more flexibility in spintronic applications [9, 10, 42].

3.3. Transport measurement and the field-driven evolution of the spin textures

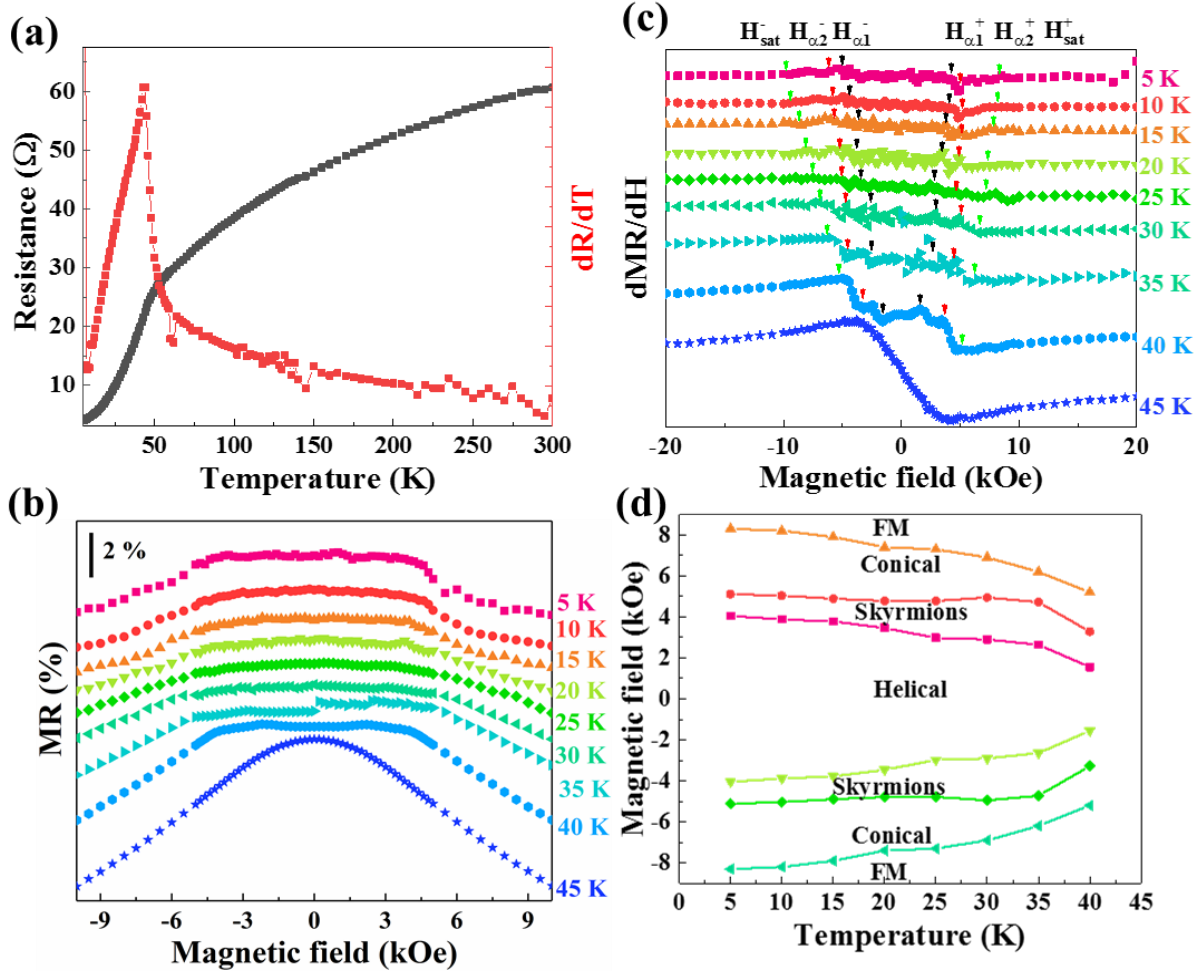

Figure 4. Electrical transport properties of the FLA-treated 30-nm-thick Mn layer on Si(100) substrate. (a) Temperature-dependent resistance (black squares) and calculated dR/dT (red squares). The peak of dR/dT indicates the position of the Curie temperature. (b) Magnetic-field-dependent magnetoresistance at various temperatures. The anomalous plateau disappears at 45 K (above the Curie temperature). (c) Calculated derivative dR/dT from data in panel (b). The black, red, and green arrows stand for the position of $H_{\alpha1}$, $H_{\alpha2}$, and $H_{sat}$, respectively. (d) Magnetic phase diagram of the MnSi film with respect to temperature and magnetic field.

The in-plane electrical resistance of the MnSi film and the calculated dR/dT are shown as a function of temperature in Fig. 4(a). The increase of resistance with temperature clearly indicates the metallic behavior [43] of the MnSi film. Though there are $MnSi_{1.7}$ impurities, the resistance is not affected due to the higher resistivity and semiconducting behavior of $MnSi_{1.7}$. It seems that the electrons bypass the cylindrical $MnSi_{1.7}$ phase. Thus, the resistance only increases a little compared with pure MnSi samples [28]. The calculated dR/dT in Fig. 4(a) has a peak, indicating the Curie temperature around 43 K, in agreement with the magnetization measurement shown in Fig. 3(a). The magnetoresistance (MR), $MR = \frac{R_H - R_0}{R_0} \times 100\%$ ($R_H$: resistance under magnetic fields, $R_0$: resistance at zero field) of the MnSi film at various temperatures is shown in Fig. 4(b). There is an anomalous plateau with some kinks at low magnetic fields and below the Curie temperature, which is due to the anomalous topology of skyrmions. Above the Curie temperature, this anomalous plateau disappears [44]. Much more pronounced kink features were reported for $Fe_3Sn_2$ [8].

We can use the derivative dMR/dH from Fig. 4(b) to investigate the field-driven evolution of the spin textures in MnSi. The same approach has been used for MnSi nanowires and other systems [8, 28, 40, 45]. As shown in Fig. 4(c), values for the positive and negative transition fields are denoted by the superscripts "+" and "−", respectively. With increasing magnetic field, the helimagnetic phase transforms into skyrmions via a first-order phase transition, manifested as peaks at $H_{\alpha 1}$ due to the completely different topological properties between helimagnetic and skyrmion phase. Above the critical field $H_{\alpha 2}$, the system transforms into the conical phase. Skyrmions can be stabilized in the range between $H_{\alpha 1}$ and $H_{\alpha 2}$. $H_{sat}$ is the critical magnetic field, above which MnSi changes into the ferromagnetic state. To better observe the change of the transition fields, the magnetic phase diagram of B20-MnSi obtained by transport measurements is shown in Fig. 4(d). The magnetic skyrmions are stable in a broad field and temperature window, in agreement with the magnetization measurements.

3.4. Skyrmion behavior dependence on the thicknesses of regrown layers

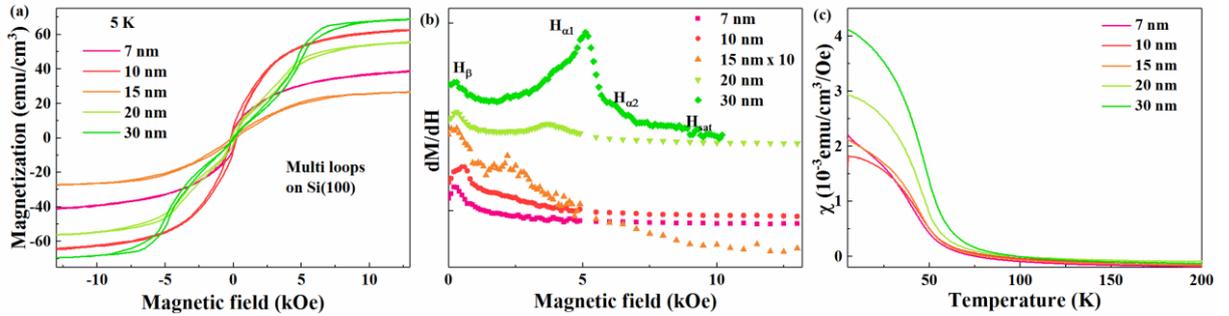

Figure 5. (a) In-plane MH curves (measured at 5 K) after FLA-treatment of Mn/Si(100) samples having different Mn layer thicknesses. The samples annealed from 20- and 30-nm-thick Mn show multi-hysteresis, indicating there is some magnetic phase transitions. (b) The calculated derivative dM/dH of the MH curves from (a). For the sample annealed from 7- and 10-nm-thick Mn, there is only one peak $H_\beta$, which is the transition from helical to concial state, since the thickness of the regrown layer MnSi is thinner than the theoretical size of MnSi skyrmions. When the Mn layer is thicker than 15 nm, an extra $H_{\alpha 1}$ peak can be detected. To better check the change of the ample annealed from 15 nm Mn, the signal is magnified ten times. (c) Magnetic susceptibility χ curves of the samples with different Mn-layer thicknesses.

To check the thickness dependence on the formation of B20-MnSi, Figure 5(a) shows the MH curves obtained from MnSi layers made by solid-state reaction of Mn films with different thicknesses. Since the theoretical size of the skyrmions in MnSi is around 18 nm, the skyrmions can be detected only when the grain size of MnSi is larger than 18 nm [37]. So, there is no multi-hysteresis, when the deposited Mn is too thin. With increasing the Mn thickness to 15 nm, first traces of the multi-hysteresis originating from the transitions of different magnetic structures can be observed, indicating the existence of magnetic skyrmions in the prepared B20-MnSi films [44]. The MnSi films prepared by ms-range solid-state reaction of 20- and 30-nm-thick Mn layers on Si(100) show obvious multi-hysteresis.

Figure 5(b) shows the calculated derivatives dM/dH. All these curves have a peak $H_\beta$ at around 300 Oe, which is the transition from the helical to the skyrmionic phase. For samples with 7 and 10 nm Mn, the system changes from helical to conical due to the smaller size of the grains [46, 47]. For thicker films, the neighbor phase is magnetic skyrmion. Since the signal of the sample annealed from 15 nm Mn is very weak compared with other samples, the dM/dH curve is multiplied by 10 to observe the change clearly. Moreover, for the sample annealed from 15, 20 and 30 nm Mn, there is also another peak $H_{\alpha 1}$, where the helical phase completely changes into the skyrmion phase. The peak $H_{\alpha 2}$ is the position, where the metastable skyrmions begin to change into the conical structure. Skyrmions are stabilized in the region from $H_{\alpha 1}$ to $H_{\alpha 2}$. The magnetic state finally changes into ferromagnetism at $H_{sat}$. The magnetic susceptibility of these samples is shown in Figure 5(c), which shows a similar transition temperature below 50 K and a different magnetic susceptibility due to the different concentrations of the MnSi phase.

3.5. Discussion

The formation of magnetic skyrmions in our MnSi films on Si(100) is indicated by magnetic and magneto-transport measurements. Since the magnetic phase diagram is similar to that of films on Si(111), the formation mechanism of magnetic skyrmions should be the Dzyaloshinskii-Moriya interaction arising from this specific B20 crystal structure. However, atom doping or the second phase could affect the lattice parameters or the anisotropy for stabilizing skyrmions. The film thickness, and hence, the grain size, can also affect the skyrmion behavior. As we show in Figure 5, if the film thickness is too thin, magnetic skyrmions can not be stabilized. We also found that with increasing the fraction of the $MnSi_{1.7}$ secondary phase, the magnetic field for stabilizing skyrmions becomes smaller. Thus, it might provide a possibility to adjust the skyrmion stabilization window by tuning the content of the second phase.

**4. Conclusions**

In summary, thin films with the B20-MnSi phase on Si(100) substrates are fabricated for the first time. The nucleation of B20-MnSi is believed to be triggered by the fast solid-state phase reaction between Mn and Si via ms-range flash-lamp annealing. Compared with the corresponding bulk material, the B20-MnSi films made by ms-range annealing show an increased Curie temperature of around 43 K. The magnetic and transport measurements reveal that skyrmions in B20-MnSi on (100) Si made by sub-seconds solid-state reaction are stable within much broader field and temperature windows than bulk MnSi. The parasitic MnSi$_{1.7}$ phase can be further minimized or eliminated by optimizing the annealing conditions, the quality of the deposited Mn film, and its interface with the Si substrate. Our work demonstrates a promising option for the fabrication of B20-type transition metal silicides for integrated and/or hybrid spintronic applications by using Si(100) wafers, which are more preferable for industry applications.

**Author contribution statement**

Y. Yuan and Z. Li initiated and designed experiments. S. Zhou, L. Rebohle and Z. Li wrote the manuscript from integrating input data and analyses provided from all the authors; V. Begeza, S. Prucnal and Z. Li synthesized samples and performed Raman measurements; T. Naumann and O. Hellwig performed X-ray powder diffraction; Authors acknowledge R. Aniol for TEM specimen preparation; R. Hübner performed TEM measurments; Z. Li performed magnetic and transport measurements. S. Zhou, K. Nielsch and M. Helm supervised the research.

**Funding**

The author Z. Li (File No. 201707040077) acknowledges the financial support by the China Scholarship Council. Furthermore, the use of the HZDR Ion Beam Center TEM facilities and the funding of TEM Talos by the German Federal Ministry of Education of Research (BMBF), Grant No. 03SF0451, in the framework of HEMCP are acknowledged. The work is also partially supported by the German Research Foundation (DFG, ZH 225/6-1).

**Declaration of Competing Interest**

The authors declare that they have no known competing financial interests or personal relationships that could have appeared to influence the work reported in this article.